\documentclass[11pt]{JHEP3}

\usepackage{amssymb,amsmath}
\usepackage{graphics}
\usepackage{epsfig}
\usepackage{amsfonts}

\usepackage{amscd}

\def\be{\begin{eqnarray}}
\def\ee{\end{eqnarray}}
\def\s{\sigma}
\def\g{\gamma}
\def\e{\epsilon}

\def\half{\frac{1}{2}}
\def\d{\partial}

\def\vf{\varphi}

\newcommand{\idn}{{1\relax{\kern-.35em}1}}

\def\g{\gamma}
\def\e{\epsilon}

\def\s{\sigma}

\def\gg{\mathsf g }

\preprint{ YITP-SB-08-10}

\title{On a worldsheet dual of the Gaussian matrix model}
\author{Shlomo S. Razamat\\
C.N. Yang Institute for Theoretical Physics\\
State University of New York\\
Stony Brook, NY 11794-3840, U.S.A.\\
E-mail: \email{razamat@max2.physics.sunysb.edu}}

\abstract{We discuss a modification of Gopakumar's prescription for constructing string theory duals of free gauge theories. We use this prescription to construct an unconventional worldsheet dual of the Gaussian matrix model.}


\begin{document}

\section{Introduction.}
Since the seminal work of 't Hooft \cite{'tHooft:1973jz} it has been widely believed that large $N$ $SU(N)$
gauge theories (with adjoint matter fields) should have a dual description in terms of closed strings.
Any correlator $M$ in a gauge theory has the following expansion
\be
M=\sum_{\gg=0}^\infty\tilde F_\gg(\lambda) N^{2-2\gg},
\ee where $\lambda=g_{YM}^2N$ is the 't Hooft coupling. This expansion has the same structure as a closed string
perturbation theory with $g_s \sim 1/N$. Of course by now we have many examples of this duality, the $AdS/CFT$ correspondence \cite{Maldacena:1997re} and the geometric transition \cite{Gopakumar:1998ki,Ooguri:2002gx} being the most widely studied in recent years. 
 However, an interesting question, still remaining open, is how to build a concrete string theory
dual for a given field theory.

In principle, a way to construct a string dual for a given theory is to know the dual of a free field theory and then to turn on
non-zero couplings. In case of gauge theories in $D<1$ this program can be pursued in some sense. The  free (Gaussian) $D=0$ matrix model is related
to a $c=-2$ matter coupled to Liouville theory \cite{Gross,Distler:1989ax}, or
equivalently to the $2d$ topological gravity \cite{Witten:1989ig,Witten:1990hr,Dijkgraaf:1990nc,Verlinde:1990ku}. The other critical matrix models can be obtained by turning on certain perturbations (see \cite{matrix_reviews,1d_review} for reviews on the subject). However, in dimensions higher than $1$ this kind of relation has not been constructed yet.\footnote{See for
instance \cite{FreeFields,StringBits,Karch,Bonelli:2004ve,Akhmedov} for different
approaches to the problem.}
 
  A very interesting prescription for constructing a stringy dual 
of a free field theory was proposed by Rajesh Gopakumar \cite{Gopakumar1,Gopakumar3} (the proposal was further investigated in \cite{Furuuchi:2005qm,Aharony:2006th,David:2006qc,Aharony:2007fs,Yaakov:2006ce}). The basic ingredient for building the duality
in Gopakumar's proposal is the well known isomorphism \cite{Strebel} between the moduli space of Riemann surfaces and the space of metric graphs. This 
isomorphism has already appeared in many string theory contexts,  like string field theory \cite{zwiebach} and different matrix models (the Penner  model \cite{Penner} and the Kontsevich
model \cite{Kontsevich:1992ti} are the standard examples, see also \cite{Mukhi:2003sz} for a recent review). 

The basic idea of Gopakumar's prescription is to think of every Feynman graph contributing to a given correlator as a metric graph, with the metric given by 
the Schwinger parameters. Then, the integration over the  Schwinger parameters is mapped to an integration over the moduli space of Riemann surfaces and
the integrand is interpreted as a worldsheet $CFT$ correlator. 

Using Gopakumar's prescription one can compute several simple correlators \cite{Aharony:2006th,David:2006qc}. These computations give rise to some interesting questions about the prescription. Namely, there are global symmetries of the gauge theory not realized globally on the
worldsheet and some of the $CFT$ correlators localize on sub-manifolds of the moduli space of
Riemann surfaces. These issues can be traced to choosing the metric on the Feynman graphs to be 
defined through the Schwinger parameters.
An important question  is whether there are different, and yet
 interesting, choices of the metric.

 In this paper we investigate a certain change of the 
metric assignment to the Feynman graphs, a choice which does not involve the use of Schwinger 
parameters. We will see in what follows that this choice of the metric ties Gopakumar's prescription closer to the much
studied cases of the matrix models. In particular we will make an attempt to explicitly construct a worldsheet dual of the Gaussian matrix model.

The rest of the paper is organized as follows. In section \ref{gop} we discuss the different issues arising from Gopakumar's prescription. In section
\ref{presc} we introduce an alternative prescription to define a metric on the Feynman graphs. In section \ref{string}  a worldsheet model for the Gaussian matrix model is built  based on the alternative prescription. In section \ref{4d} we discuss our results.

\section{Gopakumar's prescription with Schwinger parameters.}\label{gop}
Let us begin by briefly reviewing the essentials of Gopakumar's prescription \cite{Gopakumar3}
 (a detailed review of the prescription can be also found in \cite{Aharony:2006th}).
For simplicity we discuss a free gauge theory in $D$ dimensions containing a single scalar field $Q$ in
the adjoint representation of the gauge group. A generic
correlator of gauge invariant operators in this theory has the following form (again for simplicity
we discuss only single trace operators)
\be
\langle\prod_{k=1}^s TrQ^{J_k}(x_k)\rangle .
\ee The above correlator is computed by the different Wick contraction in the free theory.
Each Feynman diagram is simply a product of free propagators 
\be
\prod_{i, j=1}^s P(|x_i-x_j|)^{M_{ij}}.
\ee Here $P(|x_i-x_j|)$ is the free propagator and $M_{ij}$ is the number of the different Wick contraction of operators $i$ and $j$. 
Note that in $D>1$ the product is also restricted to $i\neq j$ by normal ordering. However,
in matrix models  the self contractions are allowed and this restriction is 
lifted. A particular way to transform each Feynman graph to a metric graph utilized by Gopakumar
 \cite{Gopakumar3} is to endow each propagator with a length parameter by representing the propagators in the following way
\be
P(|x_i-x_j|)=\int_0^\infty d\s\;e^{-\frac{1}{P(|x_i-x_j|)}\s},
\ee which is simply a variant of the Schwinger parametrization of the propagator.\footnote{Of course there are many similar choices of Schwinger parametrization. The original 
prescription of \cite{Gopakumar3} was written in the momentum space.} Then, a contribution from a specific Feynman graph  to a correlator is given by
\be\label{prop1}
\prod_{i, j=1}^s P(|x_i-x_j|)^{M_{ij}}\sim \int_0^\infty \prod_{i,j=1}^s\prod_k d \s_{ij}^{(k)}
 e^{-\frac{1}{P(|x_i-x_j|)}\s_{ij}^{(k)}},
\ee where $k$ counts the different propagators connecting given two operators and $M_{ij}$ is the
number of these propagators. An important step
in the prescription is to transform the above integral over the Schwinger parameters $\s_{ij}^{(k)}$ to
an integral over a reduced set of parameters  $\bar\s_{ij}^{(\bar k)}$. The reduced set of propagators is defined by gluing together homotopically equivalent propagators. After this
gluing the expression \eqref{prop1} can be rewritten as 
\be\label{prop2}
\int_0^\infty \prod_{i,j=1}^s\prod_{\bar k} d\bar \s_{ij}^{(\bar k)}
\left(\bar  \s_{ij}^{(\bar k)} \right)^{m_{ij}^{(\bar k)}-1}
 e^{-\frac{1}{P(|x_i-x_j|)}\bar \s_{ij}^{(\bar k)}}\qquad(\;\equiv{\mathcal A}\;),
\ee where $\bar k$ counts \textit{homotopically inequivalent} propagators connecting two operators.
The parameters $m_{ij}^{(\bar k)}$ are the numbers of homotopically equivalent propagators glued together, and satisfy $$\sum_{\bar k}m_{ij}^{(\bar k)}=M_{ij}.$$ After following the above steps
we have transformed a Feynman diagram to a metric graph which we will refer to as a skeleton graph. There are two parameters associated with
each edge, a Schwinger parameter $\bar \s_{ij}^{(\bar k)}$ which defines the metric on the graph, and an additional integer parameter $m_{ij}^{(\bar k)}$.  

The prescription then 
utilizes \textit{Strebel's} theorem \cite{Strebel,mulase,zvon} (see appendix \ref{primstr} for a very brief introduction to Strebel differentials) to rewrite the expression \eqref{prop2} as an integral
over the moduli space of punctured Riemann surfaces ${\mathcal M}_{\gg,s}$,
\be
{\mathcal A}=\int_{{\mathcal M}_{\gg,s}} [d\eta_i]\;{\mathcal F}(\eta_i, x_i, m_{ij}^{(\bar k)}),
\ee where $\eta_i$ are the coordinates on the moduli space.
We use this theorem as follows. The \textit{dual} of the skeleton graph is interpreted as 
the critical graph of a Strebel differential defined on a Riemann surface of genus $\gg$ (where
the genus is specified in the obvious way by the graph) with $s$ punctures.\footnote{Note
that this identification makes it clear why it is needed to go from Feynman graphs to skeleton
graphs, as the critical curves of the Strebel differentials contain only vertices of valence 
three or higher. We will discuss a generalization of this point in what follows.} Following the theorem,
a specific set of lengths of the edges of the graph, given in Gopakumar's case by
$\bar \s_{ij}^{(\bar k)}$, is translated to a particular point in the space 
${\mathcal M}_{\gg,s}\times {\mathbb R}^s_+$. One then is in position to translate the integration
over $\bar \s_{ij}^{(\bar k)}$ to integration over ${\mathcal M}_{\gg,s}\times {\mathbb R}^s_+$.
After explicitly performing the integrals over ${\mathbb R}^s_+$ the integrand, 
${\mathcal F}(\eta_i, x_i, m_{ij}^{(\bar k)})$, is interpreted as a $2d$ $CFT$ correlator defining
a putative string dual of our free gauge theory. The parameters $x_i$ 
are quantum numbers of the $CFT$ operators, and the numbers $m_{ij}^{(\bar k)}$ parametrize different
Feynman diagrams contributing to a specific correlator.  

\medskip
After defining a prescription to translate field theory correlators to $CFT$ correlators
one can try to compute some specific quantities. Although the above prescription
is very straightforward, it is quite complicated to use in practice. The main difficulty
being computing concrete Strebel differentials to explicitly make the transition 
between the integration variables. Nevertheless, some simple field theory diagrams can be explicitly translated to the string theory language
\cite{Aharony:2006th,David:2006qc,Aharony:2007fs}. In what follows we will be interested
in specific issues arising from these simple calculations. 

The first issue is the localization of some of the putative $CFT$ correlators obtained via
Gopakumar's prescription on submanifolds of ${\mathcal M}_{\gg,s}$. The reason for this can be
understood as follows. The real dimension of ${\mathcal M}_{\gg,s}\times {\mathbb R}^s_+$ is 
\be\label{edges} 6\gg-6+3s .\ee The number of real parameters coming from a skeleton graph is equal to the number of edges in that graph. A generic dual skeleton graph will
have only $3$ valent vertices. Obviously if we have vertices of a larger valence the number
of edges will be reduced (we do not have vertices of valence $2$ as the skeleton does 
not have homotopically equivalent edges). In such a graph the number of vertices, $v$, is related to the number of edges, $e$, by $3v=2e$. The number of faces of the dual graph is the number
of operators $s$. Then using the relation $2-2\gg=v+s-e$ we get the number of real parameters 
in a generic skeleton graph to be $6\gg-6+3s$, in agreement with \eqref{edges}.

 However,
any correlator might get contribution not just from generic graphs but also (or only)
from non-generic ones, having $n>3$ valent vertices in the dual graph. These diagrams
will have less parameters and thus after the change of variables via the Strebel theorem we will get integrals over submanifolds of ${\mathcal M}_{\gg,s}\times {\mathbb R}^s_+$. For particular diagrams 
the integration will be even restricted  to submanifolds of ${\mathcal M}_{\gg,s}$ 
\cite{Aharony:2006th,David:2006qc,Aharony:2007fs}.
\begin{figure}[htbp]
 \begin{center}
$\begin{tabular}{|c@{\hspace{0.2in}}c|c@{\hspace{0.2in}}c|}
\hline
\epsfig{file=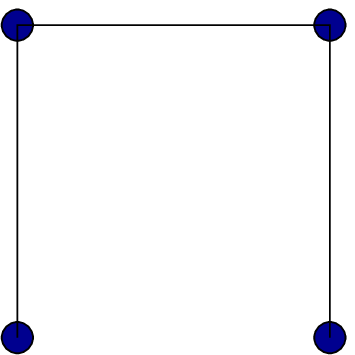,scale=0.5} & \epsfig{file=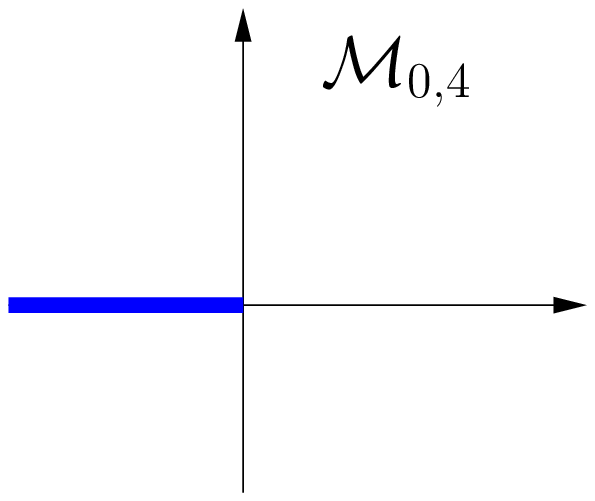,scale=0.5}\quad &\quad
\epsfig{file=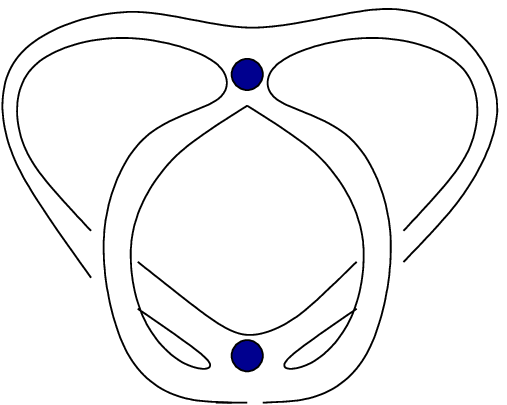,scale=0.5} & \epsfig{file=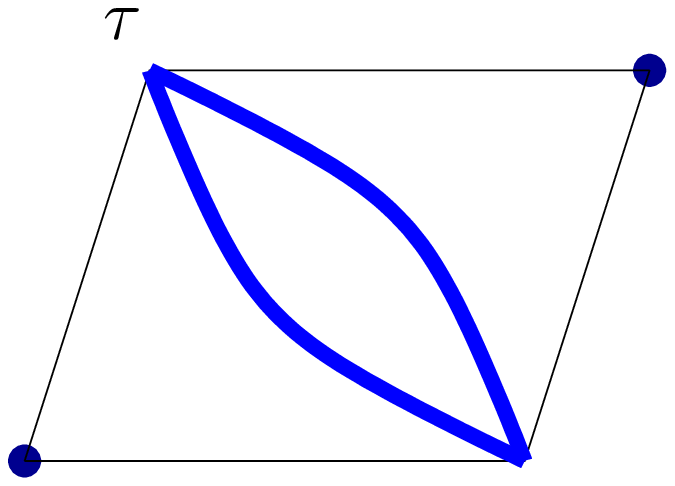,scale=0.5}
\\ [0.2cm]\hline \end{tabular}$
\caption{Examples of localizations on the moduli space. Presented are specific diagrams and
the submanifolds of ${\mathcal M}_{\gg,s}$  on which these diagrams localize. The $\Pi$ 
diagram cross ratio is restricted to the negative real values. The two point torus diagrams
without self contractions localize on co-dimension one submanifold of ${\mathcal M}_{1,2}$ (which has four real dimensions). Presented above is a set of possible positions of the second insertion, when the first one
is at the corner of the fundamental domain and the torus modulus $\tau$ is generic. These 
examples are taken from \cite{Aharony:2006th}.   } \label{locals}
\end{center}
\end{figure}
 Although we are not used to deal with $CFT$ correlators which
localize on sub-manifolds of the moduli space, one can think of mechanisms responsible
for this fact in an otherwise well behaved $CFT$s. One such mechanism was described in 
\cite{Aharony:2007fs}.
 
 It is amusing that in some cases (in the cases
depicted in figure \ref{locals}, the sphere case but not the torus one) having the localizations
might be a signal that some OPE limits are not allowed as some of the vertices do not have
Wick contraction between them on the field theory side \cite{Aharony:2007fs} (for more details on
the relation between space-time and worldsheet OPE consult \cite{Aharony:2007rq}).
 
\medskip
The second issue arising from Gopakumar's prescription is the realization
of the special conformal symmetry of the gauge theory on the worldsheet. If we take the
free theory under discussion to be scale invariant (i.e. no mass terms), the theory might
also be conformal, as it happens for gauge theories in four dimensions. The theory 
will then posses a larger symmetry which will include the dilatation transformations
and special conformal transformations. The worldsheet operators built via Gopakumar's 
prescription will have well defined dilatation properties and this symmetry will not act
on the worldsheet coordinates. This is easily seen from \eqref{prop2} for instance. The dilatation
amounts to the simultaneous rescaling of all the Schwinger parameters by a common factor. This
rescaling means that the corresponding Strebel differential is rescaled. The point on 
 ${\mathcal M}_{\gg,s}$  defined by a Strebel differential is not affected by this rescaling
and thus the only effect of it is the overall rescaling of the correlator as expected.
On the other hand, the action of the special conformal transformations is non trivial
on the worldsheet. As this action involves the space time coordinates $x_i$, it
will act differently on the different Schwinger parameters and will change the Strebel
differential corresponding to each graph, thus taking us to different points on
${\mathcal M}_{\gg,s}$. This property  does not invalidate Gopakumar's prescription, 
but makes any comparison of the obtained results to other approaches to gauge/string duality, 
like the standard $AdS/CFT$ approach \cite{Maldacena:1997re},  more difficult.

\medskip 
In the following section we will discuss how a change in the metric associated
to the Feynman graphs can affect the above listed properties.

\section{Mapping graphs to Moduli space without Schwingers.}\label{presc}

To use Strebel theorem to map gauge theory correlators to worldsheet 
expressions we had to treat Feynman diagrams as metric graphs. As we
described above R.~Gopakumar has suggested to take Schwinger parameters 
as the metric. Obviously, one has a considerable amount of freedom in rewriting
propagators in terms of Schwinger-like parameters.

As an example consider the following.
 We could define
the Schwinger parameters directly on the skeleton graph and not on the original
Feynman diagrams,
\be\label{metric2}
P(|x_i-x_j|)^{m_{ij}^{(\bar k)}}=\int_0^\infty d\s_{ij}^{(\bar k)} \exp\left(-\frac{\s_{ij}^{(\bar k)}}{P(|x_i-x_j|)^{m_{ij}^{(\bar k)}}}\right),
\ee and treat the $\s$ parameters appearing here as the metric. The putative string theory expressions  would be altered. The above prescription gives us also a natural way to avoid 
localizations. We can define a $\s$ parameter for all \textit{possible} contractions
in the skeleton graph. If some contraction does not appear in a specific Feynman graph then
the corresponding $m_{ij}^{(\bar k)}$ vanishes and the Schwinger integral integrates to $1$.
In this way we can associate a maximal number of Schwinger parameters for each 
Feynman graph and thus completely avoid the issue of localizations.\footnote{Of course the above
prescription has to be defined more carefully. For instance,  a given skeleton can be a subdiagram of different ``maximal'' graphs.} However, there are 
at least two problems even with this prescription. First, as we associate metric parameters
coupled to the space-time coordinates $x_i$ the space-time special conformal symmetry still 
remains a problem. Second, any relation between space-time and worldhseet OPE naively is lost.

The above example teaches us quite a generic lesson. Avoiding localizations means having
enough parameters on the metric graph to correspond to the real dimension of
${\mathcal M}_{\gg,s}\times{\mathbb R}^s$ and this will make the relation between the space-time
OPE and worldsheet OPE problematic. We do expect such a relation to exist from
the standard $AdS/CFT$ examples \cite{Aharony:2007rq}. Moreover, for the space-time
special conformal transformations not to act on the worldsheet coordinates the 
metric parameters on the skeleton graphs should be decoupled from the space-time 
coordinates.
 
In what follows we will describe a prescription which will cure the issue of conformal
invariance
 but it will come at the price that now \textit{all} correlators in a putative string dual to a free gauge theory  will localize on
 a discrete set of points in ${\mathcal M}_{\gg,s}$.

The way to restore the space time symmetries is to avoid direct coupling between the 
space time variables and the length parameters of the metric graphs as discussed 
above. Essentially, there is a natural 
way to associate a length parameter to an edge of a skeleton of a Feynman graph without
Schwinger parameters. As we saw in the previous section, any such edge has an integer parameter, its multiplicity $m_{ij}^{(\bar k)}$. We can define a metric on each skeleton
graph using these integer parameters. 
 Let us  consider a correlator
\be\langle\prod_{i=1}^s Tr Q^{J_i}\rangle.\ee
 For each Feynman diagram contributing to it we build a  skeleton graph as before. For each 
edge of the skeleton we specify a length given by its multiplicity, $m_{ij}^{(\bar k)}$. The 
parameters $J_i$ obviously satisfy
\be
J_i=\sum_{\bar k,\; j} m_{ij}^{(\bar k)}.
\ee Thus, we identify the parameters $J_i$ with the circumferences of the Strebel differential
poles.
 In this way we will map each diagram  to a discrete set of points in ${\mathcal M}_{\gg,s}$. Thus, any correlator 
(restricted to a given Riemann surface) in a putative string dual of the above prescription  will get contributions from a set of points in the moduli space.\footnote{Note that this prescription
retains a very weak relation between the space-time and worldsheet OPEs. Namely, as in Gopakumar's
original prescription, an absense of some OPE limit  in space-time will imply some
restrictions on the possible contributions from different regions of the moduli space.}

\medskip
To conclude this section we comment on the important issue of turning on the interactions.
First, as we will consider correlators of operators with higher dimensions we will get contributions from
more Feynman diagrams and thus more points of the moduli space will be covered.  Essentially, any Strebel differential (up to an
unimportant over-all scaling) can be approximated by a differential with integer edge lengths
(see \cite{Cach} for thorough discussion of Strebel differentials
with integer lengths). Thus as we increase the dimensions of the operators one
should get contributions from a dense set of points on ${\mathcal M}_{\gg,s}$.

After  turning on interactions,  infinite number of points of the moduli space will contribute. However,
here it is not a-priori clear whether the covering will be dense. If the covering is dense, then the string theory
expressions will be discontinuous on the moduli space (at least for arbitrary small couplings in the gauge theory). At each order of the perturbative expansion
we will  have a finite number of points and a value of the integrand will be proportional to some power of
the coupling constants. If the covering is dense then in any vicinity of a given point infinite number of orders of the
perturbative expansion will contribute.  If the couplings are not arbitrarily small
then problems might not arise. However, we will have to understand this issue better
if we are to describe weakly coupled theories, like the limit of small coupling of
the ${\mathcal N}=4$ SYM.

\section{Toward a worldsheet theory for the Gaussian matrix model.}\label{string}

In this section we will try to gain some understanding of the prescription presented above
in the relatively well understood case of the matrix models. In particular
we will consider the free, Gaussian, model. Much is known  about the relation
of this model to string theories. In particular, it is related to the 
topological gravity \cite{Witten:1989ig} and its formulation through Liouville theory with $c=-2$ matter
\cite{Distler:1989ax,Klebanov:1990sn} (see \cite{toprev} for reviews).
Moreover, there is another matrix model, the Kontsevich model, which is also related
to the topological gravity \cite{Kontsevich:1992ti}. In a modern perspective the Gaussian model is seen to be related to the
theory living on the $ZZ$ branes \cite{McGreevy:2003kb} in the above mentioned Liouville theory and the
Kontsevich model describes the string field theory on the $FZZT$ branes \cite{Gaiotto:2003yb}. The Kontsevich
model can be obtained through a specific double scaling limit of the loop operator
expectation values in the Gaussian matrix theory \cite{Maldacena:2004sn,Hashimoto:2005bf}.\footnote{See also \cite{Okounkov1}
for yet another connection between the Kontsevich model and the Gaussian theory.}  In what follows
we will not explicitly use (or try to derive) the above connections of the free matrix model to string theory.
We will rather just follow  the prescription presented in the previous section
to explicitly construct a worldsheet dual of a free matrix model.

Let us first briefly outline the main steps of the construction appearing in this section. First 
(section \ref{toy1}), we 
will use the prescription of the previous section to construct a toy $\s$-model which  will imitate a string theory structure. 
The sum over complex structures will be implemented by a sum over 
a specific set of Strebel differentials dictated by the prescription.
 We will define an action and a set of operators for this
toy model. The action and the operators will be defined in terms of a discrete set 
of fields associated to the double poles of the Strebel differentials.
This model will be explicitly designed to reproduce the Gaussian model correlators.
In the following sections we will recast this toy model as a 
worldsheet theory with an unconventional measure on the moduli space.

\subsection{A toy $\s$-model.}\label{toy1}

We consider a gauge invariant correlator of the following form
\be\label{corr}
\langle\prod_{j=1}^sTrQ^{J_j}\rangle,
\ee computed in the Gaussian matrix model with an action given by
\be\label{Maction}
S_M=\half N Tr\,Q^2,
\ee where $Q$ is a hermitian $N\times N$ matrix.

 This correlator is computed as a sum of the different Wick contractions. It can be naturally expanded in powers of $N$.
 Every loop gives a factor of $N$,  and a propagator gives $N^{-1}$. All the  vertices  are external 
 and do not give any factors of $N$. Thus the power of $N$ accompanying  each diagram contributing to an $s$ point correltor \eqref{corr} is  $f-e=2-2\gg-s$ ($f$ is the number of loops and $e$ is the number of edges in a given diagram).  Thus, a   correlator \eqref{corr} in the free theory can be written as
\be\label{prefactor}
\sum_{i} N^{2-2\gg_i-s}\,C_i= \prod_{k=1}^s J_k\,\prod_{J=1}^\infty v_J!\,N^{-s}\sum_i \frac{N^{2-2\gg_i}}{\#(\Gamma_i)},
\ee
 where summation is over different Feynman graphs and $C_i$ is the number of Wick contractions giving each graph (for instance see \cite{Penner}). 
 $v_J$ is the number of vertices of power $J$. $\gg_i$ is th genus and the factor $\#(\Gamma_i)$ is the symmetry factor of
the $i$th Feynman graph (we identify the graphs by mapping the set of vertices and the set of edges onto themselves keeping the orientation of the vertices fixed). 

\medskip
In order to construct a two dimensional  model reproducing the field theory
result we have to do essentially two things. We have to construct an action and map the field theory operators to vertex operators in the  toy $\sigma$-model. We will have a considerable amount of freedom by trading between quantities which will go to the definition
of the vertex operators and to the action. We first construct a model with the simplest
action and then use the above mentioned freedom to rewrite the expressions in a more familiar way. 

Let us first remind ourselves the logic
 of our prescription. We take a correlator of the form  \eqref{corr} and construct
all the possible Feynman diagrams. For each diagram we consider a dual of its skeleton. This 
skeleton graph has the multiplicities of its edges as the metric of those edges. Finally, due
to Strebel theorem there is a unique quadratic differential with this metric graph being its critical graph.
Thus, in order to reproduce an $s$ point correlator restricted to genus $\gg$ Riemann surface we
will have to consider all the possible Strebel differentials with integer edges on the genus $\gg$ surface with $s$ marked points,
and the residues of the poles being the parameters $J_i$. We will write an expression doing the 
job for us and then comment on the nature of different terms.
 A correlator on surface of genus $\gg$ in a free matrix model is reproduced by the following 
toy $\s$-model
\be\label{stringP}
\langle\prod_{k=1}^s\hat {\mathcal O}_{J_k}\rangle_\gg=N^{2-2\gg}\left[\sum_{\vf\in\mathcal S_{\gg,s}}\frac{1}{\Gamma(\vf)}\right]\; 
\left[\prod_{k\in P(\vf)} \int_0^1dX_k\right]\,
 e^{-2\pi i\sum_{k\in P(\vf)} X_k p_k(\vf)}\prod_{i=1}^s\hat {\mathcal O}_{J_i}.
\ee The set ${\mathcal S_{\gg,s}}$ is the set of Strebel differentials on
a genus $\gg$ Riemann surface with all the edges integer valued and $s$ double poles. $P(\vf)$ is the set of
the double poles of differential $\vf$.
 We identify the operators as
\be\label{op}
TrQ^J\,\to\; \hat{\mathcal O}_J=N^{-1}
\sum_{h=0}^{\lceil\frac{J}{2}-1\rceil}\left(\begin{array}{c}J\\h\\\end{array}\right)
(J-2h)\sum_{k\in P(\vf)} e^{2\pi i (J-2h) X_k}\equiv
\sum_{h=0}^{\lceil\frac{J}{2}-1\rceil}\left(\begin{array}{c}J\\h\\\end{array}\right)
{\mathcal O}_{J-2h}.\nonumber\\
\ee With these identifications we will reproduce exactly the free
field theory result (only the connected diagrams counted)
\be
\langle\prod_{j=1}^sTrQ^{J_j}\rangle_\gg=\langle\prod_{j=1}^s\hat{\mathcal O}_{J_j}\rangle_\gg.
\ee
Let us explain different quantities appearing above
\begin{itemize}
\item The integration over $X_k$ is introduced to equate between the insertion parameter $J$
and the circumference of the differential at puncture $k$, $p_k(\vf)$.
\item $\Gamma(\vf)$ is a symmetry factor and is computed as follows.
Note that each Feynman diagram $i$ defines several inequivalent metric graphs and thus
is described by a number of points on the moduli space.\footnote{However, each metric graph
corresponds to a unique Feynman diagram.} The inequivalent metric graphs
are obtained by assigning labels to the vertices and identifying graphs by mapping the set 
of vertices and the set of edges onto themselves keeping the orientation of the vertices fixed.
 We denote by $n_i$ the number of inequivalent metric graphs corresponding to a given Feynman diagram. Thus, to reproduce the combinatorics of \eqref{prefactor} we have
to define $\Gamma(\vf)=n_i\,\times\,\#(\Gamma_i)$.
\begin{figure}[htbp]
 \begin{center}
$\begin{tabular}{|c|c|}
\hline
\epsfig{file=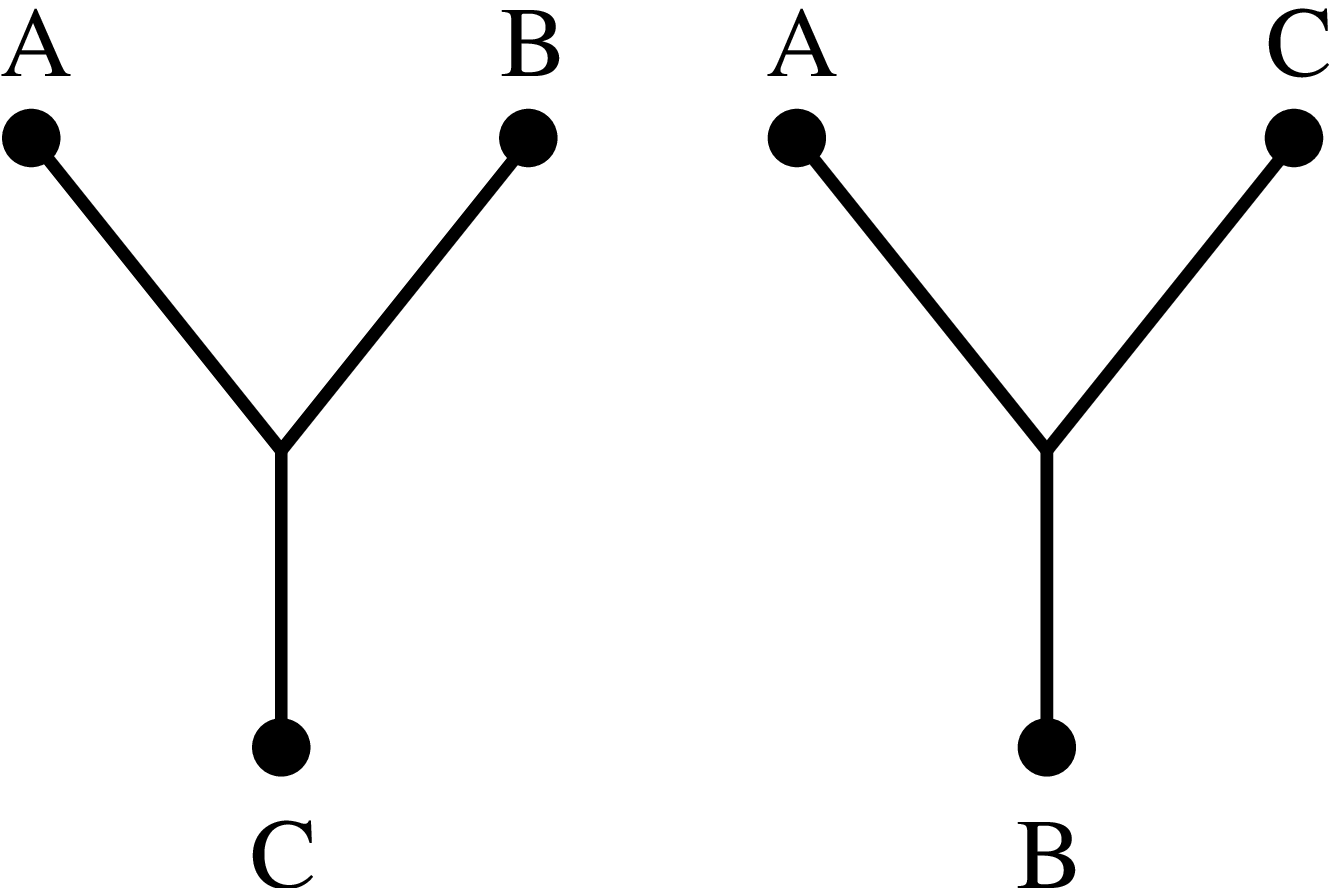,scale=0.4}  & \epsfig{file=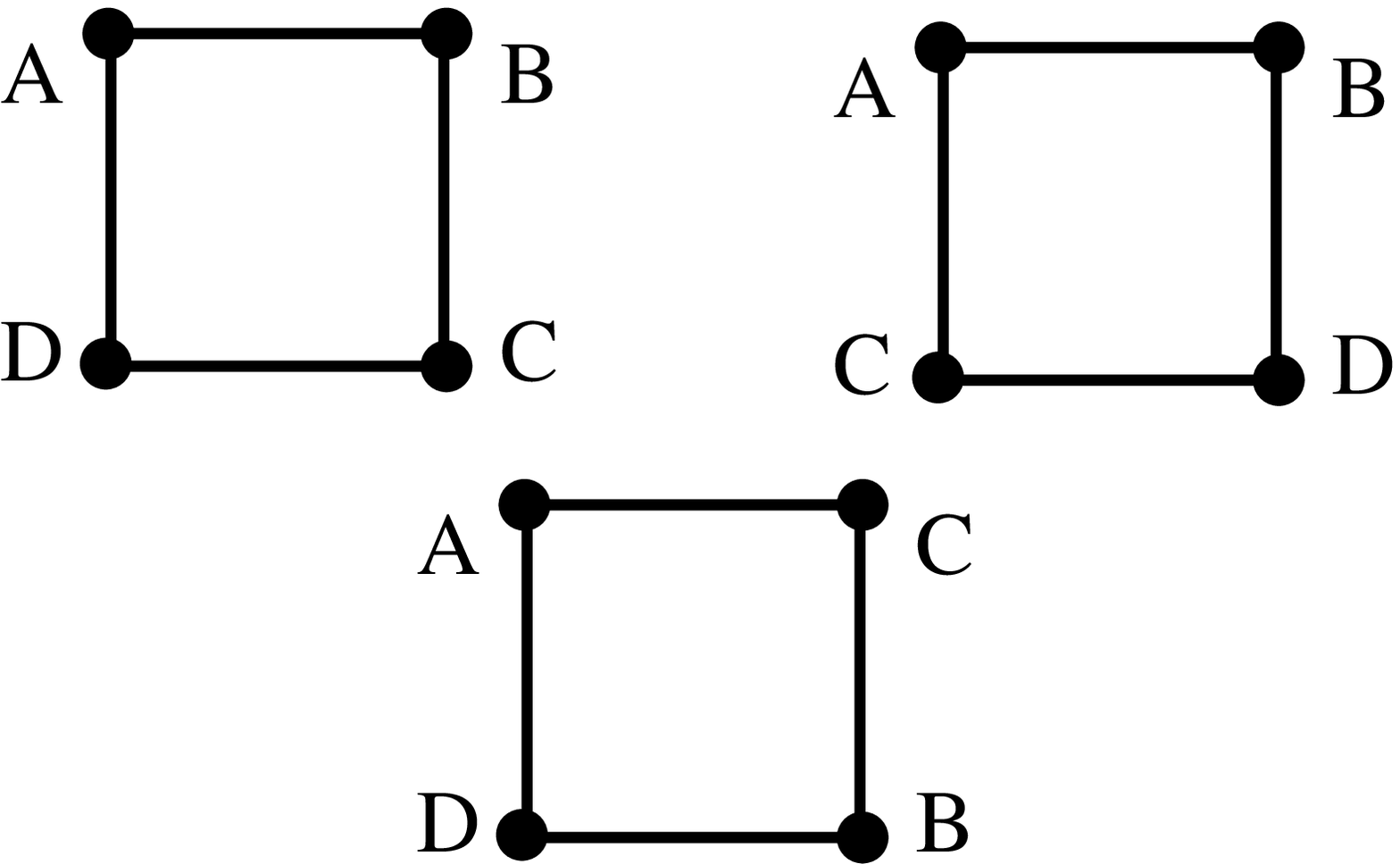,scale=0.4}
\\ [0.2cm]\hline \end{tabular}$
\caption{Depicted here are two examples of different metric graphs corresponding
to the same Feynman diagram. All the edges of the graphs are of length one. On the left we have two metric graphs 
corresponding to the unique Feynman diagram contributing to $\langle (TrQ)^3\,TrQ^3\rangle$
($n_i=2$ and $\#(\Gamma_i)=3$). On the right we have three metric graphs corresponding to the unique diagram of $\langle (TrQ^2)^4\rangle$ ($n_i=3$ and $\#(\Gamma_i)=8$).} \label{metricFeyn}
\end{center}
\end{figure}
\item The sum over Strebel differentials has to be further constrained in order
to not over-count contributions due to the CKVs. For instance, we have to take the sphere
Strebel differentials with three of the pole positions fixed to prescribed values.
\item 
The binomial coefficients appearing in the operator mapping count 
homotopycally trivial self contractions of the $TrQ^J$ operators.
\item The sum over Strebel differentials can be decomposed as follows. Any two differentials
having same critical graphs but edges of which differ by a common multiplicative factor
correspond to the same location on the moduli space. Thus, the sum over the set of Strebel differentials with integer lengths ${\mathcal S}_{\gg,s}$ decomposes
into a dense sum over the moduli space times a sum over ${\mathbb N}$ which encodes the over all integer scale of the differential. Roughly speaking
\be\label{mgk_meas}
\int'_{{\mathcal M}_{\gg,s}}\sum_{k=1}^\infty\;\longleftrightarrow\;\sum_{\mathcal S_{\gg,s}}.
\ee Here the \textit{prime} over the integral denotes that the integration is really a dense summation.
\end{itemize}

In the following sections we will rewrite the above model in a language a bit more familiar
from the usual formulations of string theories.

\subsection{Enlarging the set of differentials.}\label{punct}

In this section we will discuss a useful extension of the set of Strebel differentials
${\mathcal S}_{\gg,s}$ defined in the previous section. As we will see in the following
section this extension will allow us to rewrite our toy $\sigma$-model in an (almost)
conventional string theory language.

The extension we will introduce is basically to add to the set
${\mathcal S}_{\gg,s}$  differentials with some of
the circumferences exactly zero.
There are no Strebel differentials with zero circumferences and thus we should extend the notion
of these differentials to include such objects. Essentially, the extension is straightforward. Note
that we can add to the set of Strebel differentials limits of differentials with one of the edges going to zero length. Usually
this procedure brings two zeros together and produces a zero (with a valence being a sum of the valences of the two original zeros). However, if the edge encircles a double pole (and has both ends ending on the same vertex) then a more interesting scenario can happen. If the edge
ends on a zero of valence greater than two the result of taking the limit is a zero of smaller valence. If the zero is of valence two then the result of the limit is a regular point.
Thus, the above two limits produce well behaved Strebel differentials. Finally,
if the edge ends on a simple zero the result of the limit will be a simple pole and
the critical graph will have a single edge emanating from this point. A Strebel differential
can not have simple poles and thus we will have to add the above configuration to the differentials
we consider.\footnote{A discussion of quadratic differentials with simple poles 
appears also when considering compactifications of moduli spaces. See for instance \cite{zvon}. 
}
\begin{figure}[htbp]
 \begin{center}
 $\begin{array}{c@{\hspace{0.6in\vspace{0.3in}}}c}
\epsfig{file=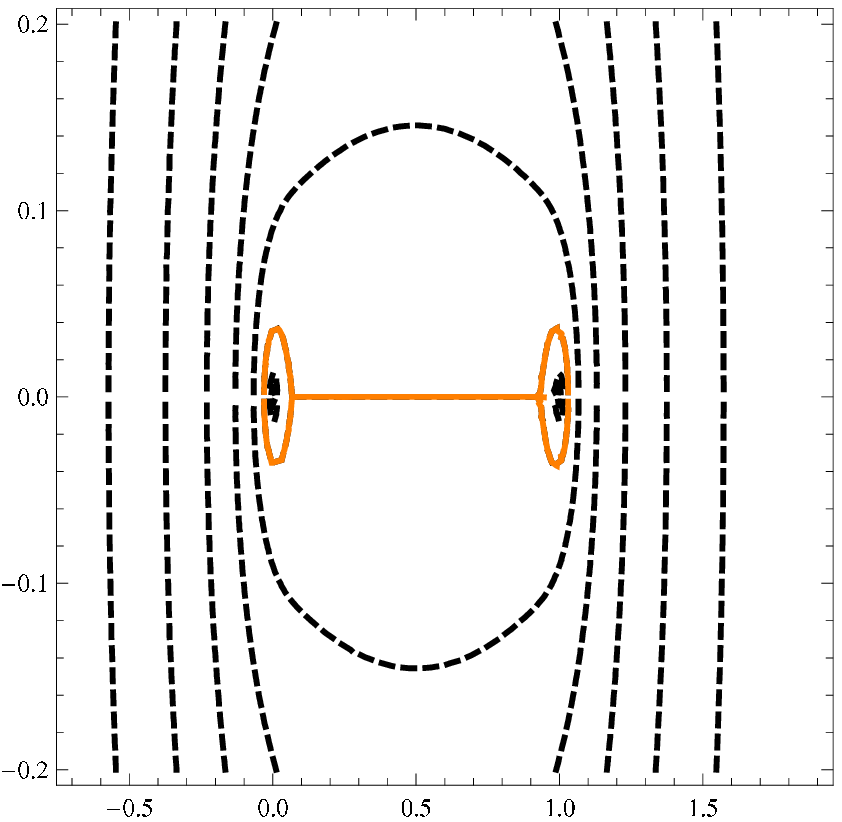,scale=0.55} &  \epsfig{file=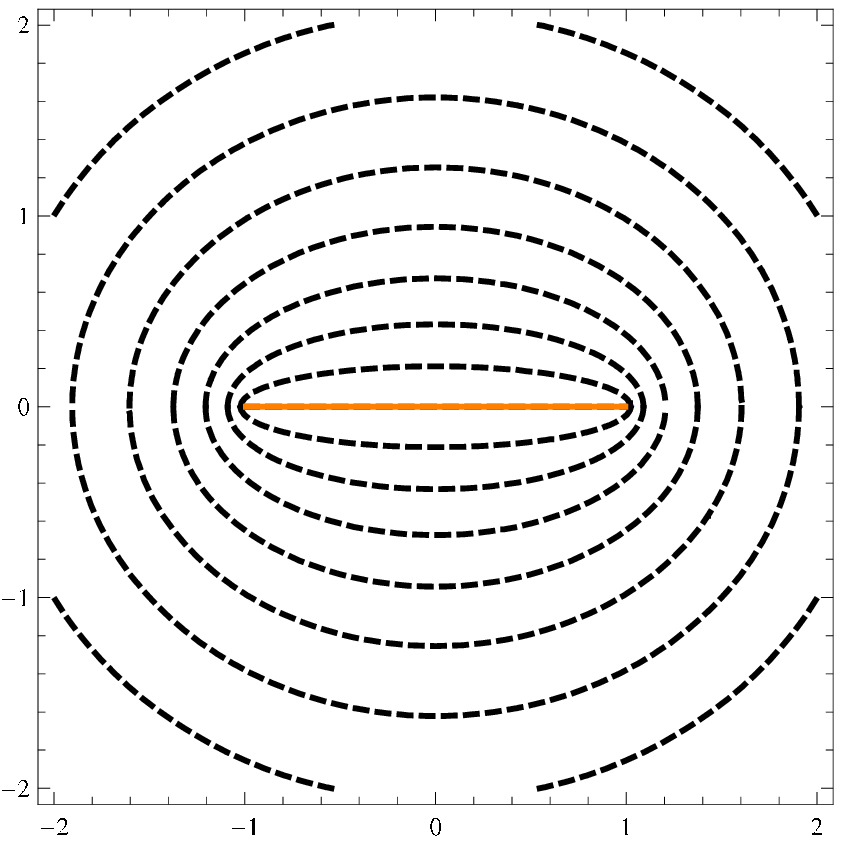,scale=0.55}
\\ [0.2cm]
\end{array}$
\caption{Two circles contracting to  zero size in the left diagram to give a critical curve  with two simple poles.} \label{PPTrQ2}
\end{center}
\end{figure} 

Consider the following two simple examples of the  differentials
that the above procedure produces. Consider the differential
\be
q(z)\,dz^2=-\frac{1}{4\pi^2}\left[a\left(\frac{dz}{z}\right)^2+b\left(\frac{dz}{z-1}\right)^2
+c\left(\frac{dz}{z(1-z)}\right)^2\right].
\ee For $ab+ac+bc<0$ the critical curve of this differential looks 
as the left curve on figure \ref{PPTrQ2}. By taking $a=b=-c$ we obtain the following differential
\be
q(z)\,dz^2=-\frac{a}{2\pi^2}\frac{1}{z(z-1)}dz^2,
\ee  and the critical curve is depicted on the right side of figure \ref{PPTrQ2}.
A second simple example is the case obtained by a different limit of the diagram depicted
in \ref{PPTrQ2}. One can take one of the circle edges to zero as well as the central edge.
This is given by $b=c=0$ and the differential is simply
\be
q(z)dz^2=-\frac{a}{4\pi^2}\frac{1}{z^2}.
\ee

Note that introducing simple poles enables us to discuss homotopically trivial 
self contractions on the same footing as all the other contractions.  Thus we will make the following definition
of a set of differentials which will be of interest to us.
We will denote the set of the Strebel differentials  with integer lengths on
genus $\gg$ surface with exactly $s$ double poles with non-vanishing circumferences and any number of double poles with vanishing circumferences
(in the sense discussed above)  by
 $\hat {\mathcal S}_{\gg,s}$. This set extends ${\mathcal S}_{\gg,s}$ by allowing differentials
with simple poles.

Let us discuss two possible metrics one can put
on the worldsheet using a Strebel differential $\vf$.
One natural choice to define a metric is
\be
ds^2=g_{z\bar z}dzd\bar z=|\vf| dz d\bar z.
\ee The curvature tensor of this metric is proportional to $\d\bar\d |\vf|$ and we can write
\be
\frac{1}{4\pi}\sqrt{g}R=-\half\sum_{k}m_k\delta^2(z-z_k,\bar z-\bar z_k),
\ee where $z_k$'s are the positions of the poles and the zeros of the differential. The
numbers $m_k$ are the multiplicities of these points. That is for a double pole $m_k=-2$,
for a simple pole $m_k=-1$, for a regular point $m_k=0$ and for a zero of valence $v$ we have
$m_k=v$. One can check the combinatorics to obtain 
\be
\frac{1}{4\pi} \int d^2z \sqrt{g}R= 2-2\gg.
\ee Another choice of metric follows directly from the extension of the set of differentials discussed above. Recall how the Strebel differentials entered into the prescription
of section \ref{presc}. These differentials had critical graphs equivalent to the dual graphs of 
the skeleton of a given Feynman diagram. However, after extending the set of differentials above,
it is sensible now to associate for each Feynman diagram itself a differential, which we
will denote by $\vf_D$. 
 This differential will have the Feynman diagram itself (note, \textit{not}
the skeleton but the diagram itself) as its critical graph. The existence and the uniqueness 
of such differential is guaranteed by the Strebel theorem and the construction above.
In appendix \ref{dual} we explicitly construct the dual $\vf_D$ for a given differential $\vf$.
 The vertices
of the critical graph of $\vf$ will become the faces of the critical graph of $\vf_D$ and vice versa. In the following table the map between 
the special points of the two differentials is summarized,
\be
\begin{tabular}{|c|c|}
\hline
$\vf$ & $\vf_D$\\
\hline\hline
$-\frac{p^2}{4\pi^2}\frac{1}{z^2}$ & $z^{p-2}$\\\hline
$z^m,\;m\geq-1$ & $-\frac{(m+2)^2}{4\pi^2}\frac{1}{z^2}$\\\hline
\end{tabular}\;.
\ee
Note that the double poles
with circumference $2$ of $\vf$ are mapped to regular points of $\vf_D$ and some regular
points of $\vf$ can be mapped to double poles of $\vf_D$ (for details consult appendix \ref{dual}). In what follows when we will
refer to a set of special points of a differential (zeros and poles) we will always consider
them as defined by the $\vf_D$ differential.
 
We can use the differential $\vf_D$ to define a metric 
\be
ds^2=(g_D)_{z\bar z}dzd\bar z=|\vf_D| dz d\bar z.
\ee The curvature computed in this metric has the following form
\be
\frac{1}{4\pi}\sqrt{g_D}R_D=-\half\sum_{k}(p_k-2)\delta^2(z-z'_k,\bar z-\bar z'_k),
\ee where $z'_k$ are the zeros, simple and double poles of $\vf_D$. In the above equation
the parameters $p_k$ are defined by the differential $\vf$ and in particular the circumference of a zero (or a simple pole) of $\vf$ is defined to vanish. 

\subsection{Constructing a worldsheet model.}\label{final}

We are now in position to rewrite our toy $\sigma$-model in a more familiar way.  Our action
\eqref{stringP} and the definition of operators \eqref{op} appear as discrete
sums over special points on the worldsheet. However, using Strebel differentials
we can write these as integrals over the worldsheet. 

First we claim that the following model is equivalent to the one defined in section
\ref{toy1}. The correlators in the model are computed using the following expression
\be\label{stringPv2}
\langle\prod_{k=1}^s{\mathcal O}_{J_k}\rangle_\gg&=&\left[\sum_{\vf\in\hat {\mathcal S}'_{\gg,s}}\frac{1}{\Gamma(\vf)}\right]\; 
\left[\prod_{k\in Z(\vf_D)\cup P(\vf_D)} \int_0^1dX_k\right]\,
 e^{S}\, \prod_{k=1}^s{\mathcal O}_{J_k}.
\ee The action is given by
\be\label{action}
S&\equiv& -2\pi i\sum_{k\in Z(\vf_D)\cup P(\vf_D)} \left[p_k(\vf)-2\right] X_k 
+\mu\sum_{k\in P(\vf_D)}\, e^{-4\pi i X_k},
\ee
 and we identify the operators as
\be\label{opv2}
TrQ^J\,\to\;{\mathcal O}_J=N^{-J/2}
J\sum_{k\in Z(\vf_D)} e^{2\pi i (J-2) X_k}.
\ee The set $P(\vf_D)$ is the set of the double poles, and
$Z(\vf_D)$ is the set of simple poles and zeros of the $\vf_D$ differential.
The parameter $\mu$ will be shortly identified. 
We see that with the above identification of the operators $TrQ^2$ is
problematic. Thus, we will remove this operator from the theory and restrict the set
 $\hat{\mathcal S}_{\gg,s}$ to differentials with no double poles of
circumference two. This set is denoted  by $\hat{\mathcal S}_{\gg,s}'$ in equation 
\eqref{stringPv2}. 
We will comment on the interpretation of the  $TrQ^2$ operator at
the end of this section.

The equivalence of \eqref{stringPv2} and \eqref{stringP} can be easily established.
Note that the integration over $X_k$ for $k\in Z(\vf_D)$ sets $p_k=J_k$ exactly as 
one obtains in the model of section \ref{toy1}.
Further, the existence of homotopically trivial self-contractions
is dealt with by allowing simple poles. Note that a simple pole of the differential $\vf$ 
implies that the corresponding Feynman diagram has a homotopically trivial self-contraction,
and thus summing over all possibilities to have simple poles counts all the different ways
to introduce this class of self contractions.
  The integration over $X_k$ for $k\in P(\vf_D)$
gives simply an overall factor of $\mu^{f}$, where $f$ is the number of faces of the Feynman
 diagram (which is equal to the number of the double poles of $\vf_D$). The overall $\mu$ and $N$ dependence of the correlator is
\be
N^{-\half\sum_{k=1}^sJ_k}\mu^{f}=N^{2-2\gg-s}\left(\frac{\mu}{N}\right)^f.
\ee Thus, by setting $\mu=N$ the model \eqref{stringPv2} is equivalent to
the model \eqref{stringP} and thus to the Gaussian matrix model.

The reason why we introduce a slightly modified version of our original simple model
is that now we can naturally rewrite it in terms of real worldsheet fields.
Using the results of the previous subsection we can write the $X$ dependent part of the action as
( denote ${\mathcal H}=Z(\vf_D)\cup P(\vf_D)$ )
\be\label{Xint}
-2\pi i\sum_{k\in {\mathcal H}} \left[p_k(\vf)-2\right]\,X_k&=&-2\pi i\int d^2z\, \sum_{k\in {\mathcal H}}\, (p_k-2)\delta^2(z-z_k,\bar z-\bar z_k)\, X(z,\bar z),\nonumber\\
&=& i \int d^2z \sqrt{g_D}R_D X(z,\bar z).
\ee Here $X(z,\bar z)$ is a function on the Riemann surface satisfying
\be
X(z_k,\bar z_k)=X_k.
\ee  In order to define more concretely how to lift the set of $X_k$s to a field
on the worldsheet we will have to define the functional integration measure of this field.
To do so we first write the vertex operators as functions of this field
\be\label{opZCCM2}
TrQ^{J}\quad\to\quad {\mathcal O}_J=
 N^{-J/2}\,
\int d^2z\,f_J(g,g_D)\,e^{2\pi i(J-2)X(z,\bar z)}.\nonumber
\ee  Here $f_J(g,g_D)$ is a yet undetermined function of the metrics.  
Next we denote the appropriate measure for the field $X$ by $[\hat{\mathcal D}X]_{g_D}$.
The functions $f_J$ and the definition of the measure should conspire so that
the functional integration will reproduce the discrete integrations of  \eqref{stringPv2}. 
There is a considerable amount of freedom of how to define these quantities. However,
one can show (the technical details can be found in appendix \ref{measure}) that the
following choice can be made. One can  define
\be 
f_{J>0}(g,g_D)=C_J\,\sqrt{g},
\ee where $C_J$'s are normalization factors (see appendix \ref{measure}).
With this choice the discrete theory is reproduced by the following functional integration
\be
[\hat{\mathcal D}X]_{g_D}=\Delta_{g_D}[{\mathcal D}X]_{g_D}\,e^{-\pi\int d^2z\,\d X\bar\d X},
\ee where $\Delta_{g_D}=\sqrt{\det\,\d\bar\d}$ and $[{\mathcal D}X]_{g_D}$ is the standard measure
for a periodic scalar. 

 Further, the $\mu$ dependent part
of the action can be written as $\hat\mu\, {\mathcal O}_0$ where 
\be\label{pun}
{\mathcal O}_0&=&\int d^2z\,\sqrt{g_D}\,e^{-4\pi iX(z,\bar z)},
\ee and we will refer to this operator as the ``puncture'' operator.  The parameter 
$\hat\mu$ is defined as \be\hat\mu\equiv C_0\,\mu.\ee Here $C_0$ is a normalization constant (see appendix \ref{measure}). Note that the definition
of the functions $f_J$ for the puncture operator and for the other operators is different. The technical reason for this, as explained in appendix \ref{measure},  is the fact that the puncture operator has to be inserted on the faces of the Feynman diagram and the other operators are inserted at the vertices of the diagram.

Let us summarize our results. The correlators (only connected diagrams counted)
of the Gaussian matrix theory are reproduced by the following model. The correlators
are given by
\be\label{stringff}
\langle\prod_{k=1}^s{\mathcal O}_{J_k}\rangle_\gg=\sum_{\vf\in\hat{\mathcal S}'_{\gg,s}}\frac{\Delta_{g_D}}{\Gamma(\vf)}\; 
 \int[{\mathcal D} X]_{g_D}\,
 e^{-S_X(\hat\mu)}\prod_{k=1}^s{\mathcal O}_{J_k}.
\ee The action is 
\be
\boxed{S_X(\hat\mu)=\pi\int d^2z \d X\bar \d X-i\int d^2 z\sqrt{g_D}\,R_D\,X-\hat\mu\,\int d^2z\sqrt{g_D} e^{-4\pi i X}}\;,
\ee and the operators are
\be
\begin{tabular}{|l|l|}
\hline
${\mathcal O}_0=\int d^2z\,\sqrt{g_D}\,e^{-4\pi iX(z,\bar z)}$&
${\mathcal O}_{J\neq 0,2}=N^{-J/2}\,C_J\,\int d^2z\,\sqrt{g}\,e^{2\pi i(J-2)X(z,\bar z)}$%&
\\\hline
\end{tabular}\;.
\ee

Note that the above theory bares some resemblance to the Itzhaki-McGreevy string theory \cite{Itzhaki:2004te}. In particular  it is tempting to identify $-2\pi X$ with the time-like Liouville field of that model (which has $c=25$).

To actually prove that the model \eqref{stringff} is equivalent to some kind of string theory
we have to understand better the measure on the moduli space.
The measure on the moduli space is unconventional and is given by
\be\label{measureF}
\sum_{\hat{\mathcal S}'_{\gg,s}}\frac{\Delta_{g_D}}{\Gamma(\vf)}.
\ee We did not start by defining our model as an integral over metrics, but it is plausible to guess that the above measure
should consist of the integration over the usual moduli space of Riemann
surfaces together with some ghost system and an additional matter field(s). It is very interesting to establish whether these
can be formulated in terms of a conventional $bc$ ghost system and a $c=1$ matter field, which  should be the real Liouville mode 
in this model, the one coming from the broken Weyl invariance. A remnant
of this Liouville field can be seen in \eqref{mgk_meas} in terms of a discrete rescaling of the Strebel differentials and thus also
a rescaling of the $g$ metric. In this context, let us comment on the nature of the $TrQ^2$ operator. The fact that we fail
to describe this operator on the worldsheet is yet another sign for the need to
introduce additional structure to the model. It might be the case that the ${\mathcal O}_2$
operator should be defined in terms of the additional worldsheet fields and be independent of the field $X$.
 We leave the investigation of these issues to future research.

\section{Discussion.}\label{4d}

In this paper we have introduced a prescription for constructing
worldsheet duals of free field theories. This prescription is a variant
of the prescription suggested by R.~Gopakumar \cite{Gopakumar3}. 
Using this prescription we have constructed an unconventional worldsheet dual of the free matrix model. 

There are many questions which deserve further investigation.
First of all, the main issue left open is a 
better understanding of the moduli space measure. In particular one can wonder
whether a string theory in an unconventional gauge might give the measure \eqref{measureF}.
Making sense of the measure on the moduli space is crucial for making any 
connection with the familiar formulations of string theories.

Another important direction for further research is to generalize the discussion
in section \ref{string} to higher dimensions and to theories with multiple fields.
The similarity of our model to the Itzhaki-McGreevy string, which was proposed in \cite{Itzhaki:2004te} as a worldsheet
dual of the quantum mechanics of a gauged matrix harmonic oscillator,\footnote{See also \cite{Berenstein:2004kk} for a discussion of harmonic oscillator quantum mechanics 
in the context of gauge/gravity duality.} suggests that 
a better understanding of the moduli space measure might shed some light on the $D=1$ case.
Essentially, it is tempting to conjecture that the unconventional worldsheet dual of the Gaussian
 matrix model obtained in section \ref{string} is really computing a certain class of harmonic oscillator expectation values, and the full string theory extension of this model is  
the dual string theory of the harmonic oscillator quantum mechanics.

There are at least two new issues with going to $D\geq 2$. First, the propagators
are not trivial any more, i.e. they depend on the space-time coordinates $x_\mu$. One can
try and use the Fourier tricks of section \ref{string} to deal with this complication. That is, we can introduce additional $D$ fields $Y_\mu$ on the worldsheet. 
For each point of ${\mathcal S}_{\gg,s}$ the action for these fields can be defined
as a Fourier transform of the corresponding Feynman diagram. The vertex operators thus
will include an additional exponential $\exp(-ix_\mu\,Y^\mu(z_k,\bar z_k))$. The challenge then is to write a continuous (and local) action on the worldsheet for the fields $Y_\mu$.      

Another important new issue with going to $D\geq2$ is the absence of self-contractions 
in these theories. As the position space propagators diverge when the separation is
taken to zero, we have to remove all the self contractions with some kind of normal ordering procedure. 
Removing self contractions implies that we will have less Feynman diagrams and  cover
less points on the moduli space. Essentially, the self contractions can be divided into two
classes, homotopically trivial and non trivial ones. We saw in section \ref{final} that the
homotopically trivial self contractions can be naturally taken into account by the puncture
operator. Thus, taking away this class of self  contractions will probably amount
to reconsidering the puncture operator. Taking away the homotopically non trivial 
self contractions has a more geometric meaning. For instance, the points contributing to any
two point correlator on the torus will localize on a submanifold of ${\mathcal M}_{\gg=1,2}$
exactly as it happens in Gopakumar's prescription 
\cite{Aharony:2006th}. It will be very interesting to understand better the role
played by self contractions in this type of constructions. 

\acknowledgments 

I am grateful to O.~Bergman, R.~Gopakumar, A.~Pakman, and L.~Rastelli for useful discussions and comments on the manuscript. I would  especially like to thank O.~Aharony and Z.~Komargodski for numerous discussions and comments during different stages of this project. This work was supported in part by the National Science Foundation, grant 0653342.

\appendix

\section{ A short primer on Strebel differentials.}\label{primstr}

A quadratic differential is the following object,
\be
q=\vf(z)\, dz^2,
\ee where $\vf$ is a meromorphic function on a given Riemann surface. This differential is defined to have the following property under a holomorphic reparametrization of the worldsheet
 $z\to z'(z)$,
\be
\vf(z)dz^2=\vf'(z')(dz')^2.
\ee Using quadratic differentials one can define a length for a line element through
\be\label{strmet}
dl=\sqrt{\vf}\,dz.
\ee Note that this length is in general a complex number.
It is useful to define the notions of horizontal and vertical curves of the differential.
Given a curve $\g(t)$ on the Riemann surface we say that it is horizontal if 
\be
\vf(\g(t))\left(\frac{d\g}{d t}\right)^2>0,
\ee and vertical if the opposite inequality holds. Note that the length
of the horizontal curves computed using \eqref{strmet} is real. By convention we will discuss 
the horizontal curves in what follows. A horizontal curve can either be closed or
end on a zero or a pole.  
The set of all non-closed horizontal curves of a quadratic differential is called the critical
curve of the differential. We restrict to quadratic differentials critical curve
of which is compact.
\begin{figure}[htbp]
 \begin{center}
 $\begin{array}{c@{\hspace{0.05in}}c@{\hspace{0.05in}}c@{\hspace{0.05in}}c}
\epsfig{file=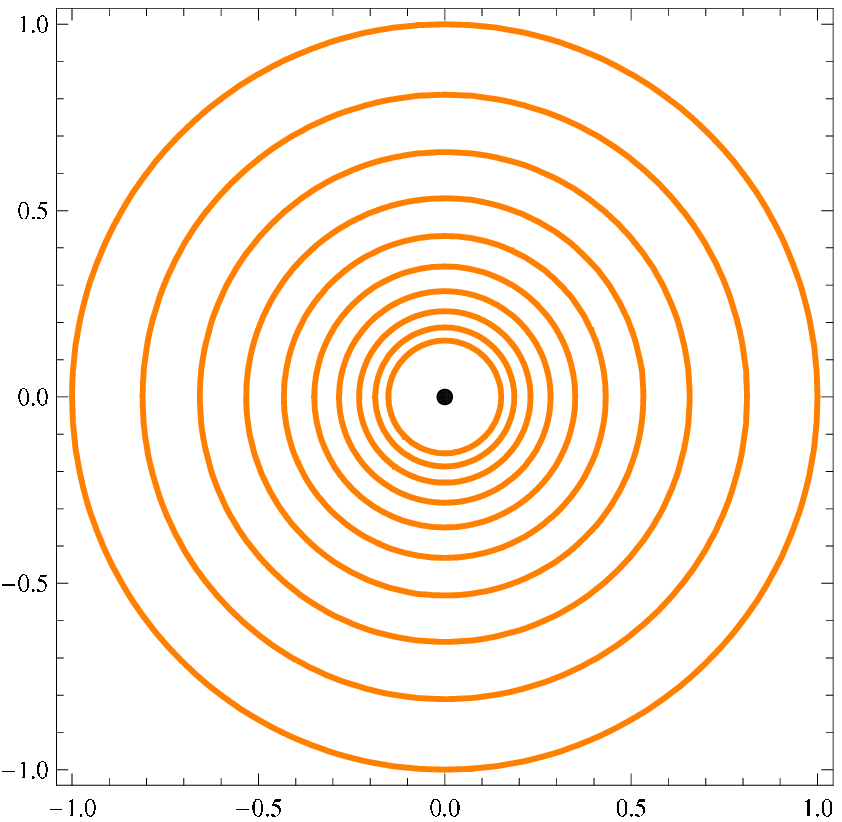,scale=0.4} & \epsfig{file=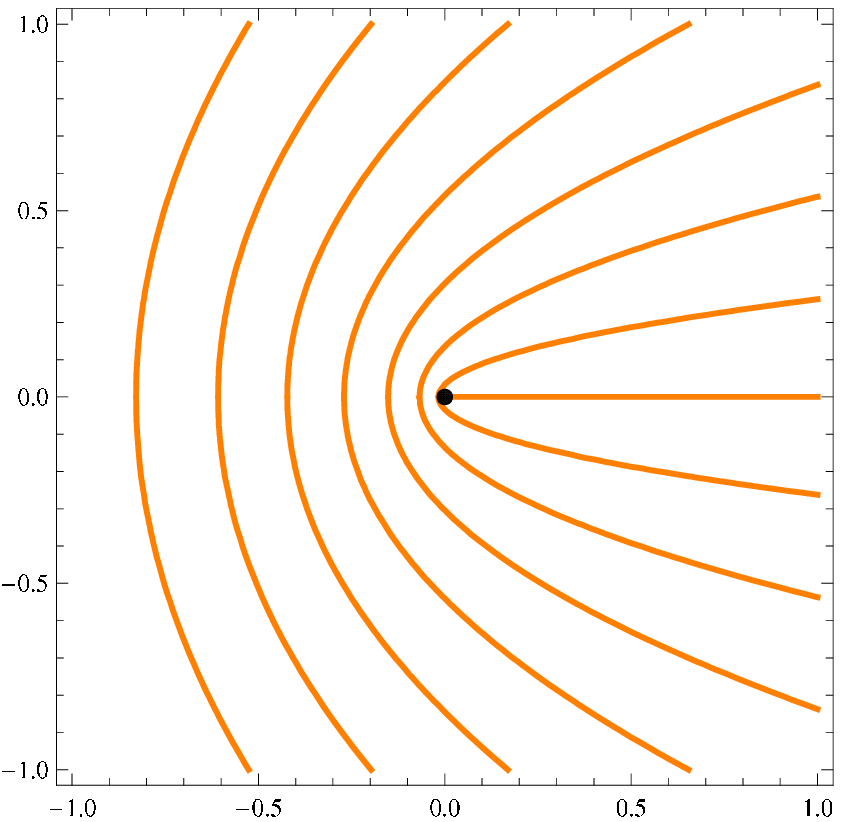,scale=0.4}&\epsfig{file=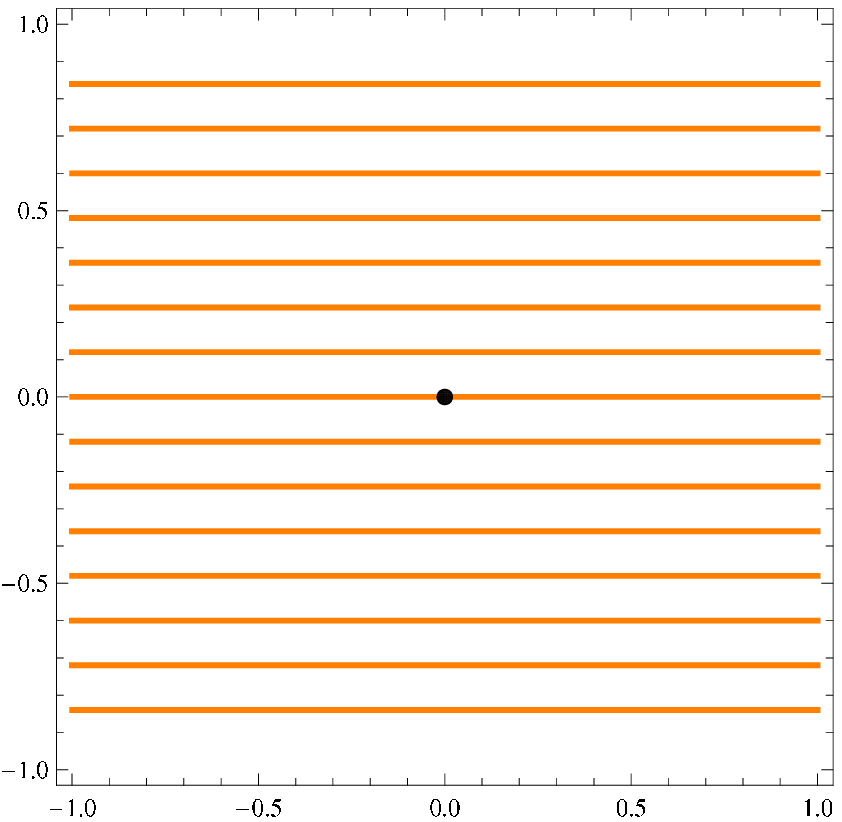,scale=0.4} &\epsfig{file=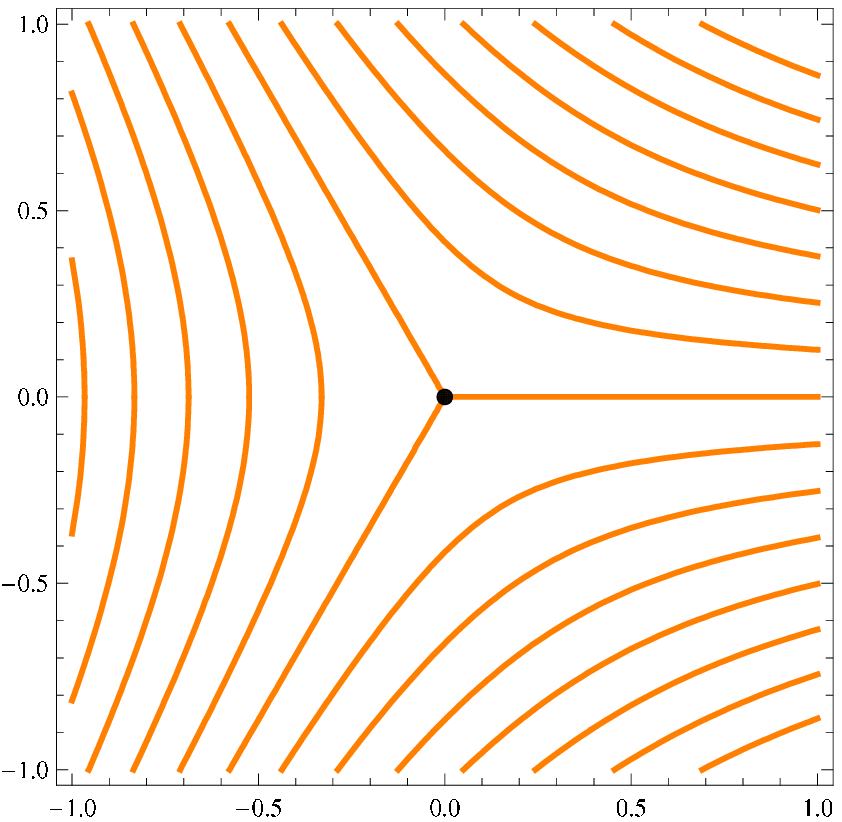,scale=0.4}  
\\ [0.2cm]
\end{array}$
\caption{Behavior of the horizontal curves of a quadratic differentials in the vicinity of different points. From left to right we have
a double pole (with a negative coefficient), a simple pole, a regular point and a simple zero.} \label{strebels}
\end{center}
\end{figure}
If a quadratic differential has at most double poles (with negative coefficients)
 then the critical curve divides the Riemann surface into ring domains. The vertices of the critical curve of such a differential are the zeros and the simple poles of the differential.
The following theorem due to K.~Strebel holds,

\noindent\textit{
Given a Riemann surface with $s$ marked points and $s$ positive numbers $p_k$ associated
to those points, there is a unique quadratic differential with double poles as its only
singularities such that:
\begin{itemize}
\item It has exactly $s$ double poles located at the marked points
\item The residues of the double poles are the numbers $p_k$
\item The Riemann surface is a union of $s$ disc domains defined by the marked points. 
\end{itemize}
}

We refer to a differential which satisfies the properties above as a Strebel differential.
Note that from this theorem follows that there is a unique  Strebel differential for each point of 
 ${\mathcal M}_{\gg,s}\times{\mathbb R}_+^s$. Further, this also gives us a natural
isomorphism between the space ${\mathcal M}_{\gg,s}\times{\mathbb R}_+^s$ and the space
of metric graphs with $s$ faces on genus $\gg$ surface.\footnote{In this context we define a metric graph as a connected graph with
a positive real number associated to every edge and all the vertices at least trivalent.} For each point of
${\mathcal M}_{\gg,s}\times{\mathbb R}_+^s$ we associate the critical curve of the corresponding
Strebel differential as the metric graph (metric on the graph defined through \eqref{strmet}),
and the other direction of the isomorphism can be also (less trivially) established. 

\begin{figure}[htbp]
 \begin{center}
 $\begin{array}{c@{\hspace{0.8in}}c}
\epsfig{file=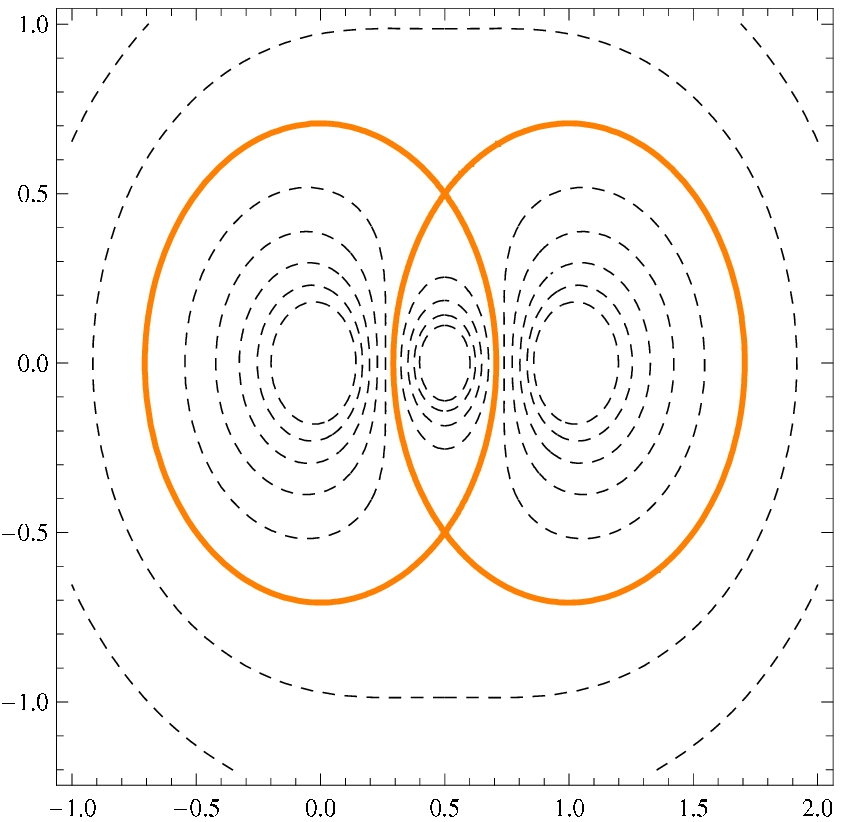,scale=0.6} & \epsfig{file=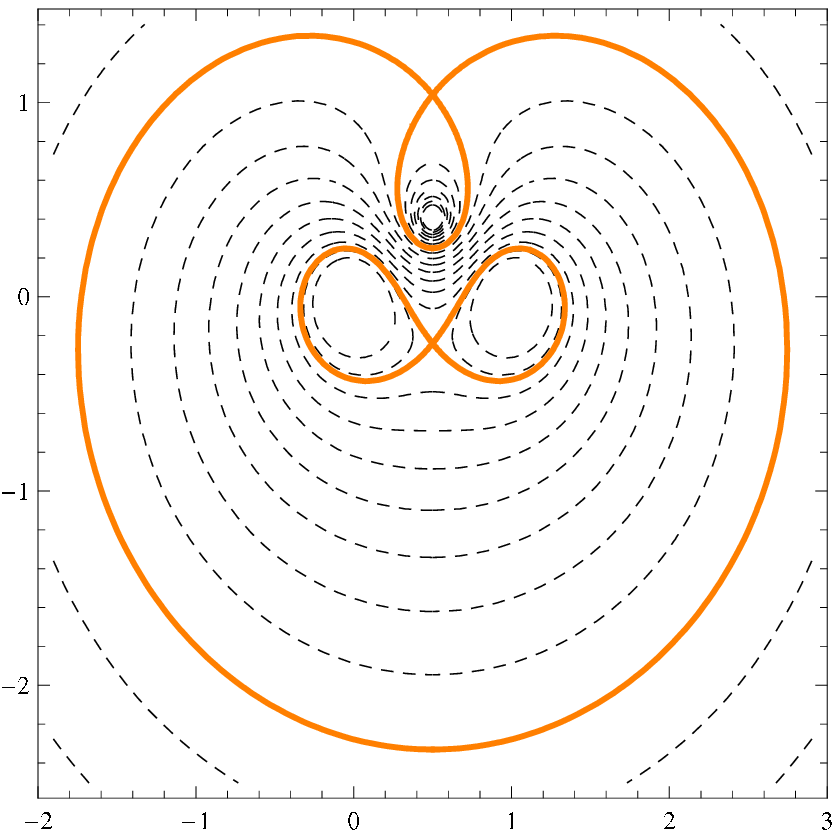,scale=0.6}
\\ [0.2cm]
\end{array}$
\caption{On the left we have an example of a Strebel differential on a sphere. It has four
double  poles located at $z=0,1,\infty$ and $z=\half$. We see that the sphere is decomposed
into four disc domains. On the right we have a differential which is not Strebel. 
It differs from the left one by taking  the fourth insertion from $z=\half$ to 
$z=\half+0.4\,i$. Note that now we have one additional ring domain. When one will
compute the distance between the two zeros of the differential (the two vertices of the critical curve) it will be a complex number.} \label{square}
\end{center}
\end{figure}

When explicitly trying to find a Strebel differential for a given Riemann surface and a given set
 of residues the first two conditions above can be easily satisfied. The third condition is
however a very non-trivial one. Essentially, it can be rephrased as the demand that all the 
distances between the zeros of the differential computed in Strebel metric \eqref{strmet} should
 be real. Computing these distances will give constraints on the parameters of the differential. 
 Usually these constraints will be expressed through elliptic integrals, which are difficult to solve.
  In figure \ref{square} an example of a Strebel differential is depicted.

\section{The dual Strebel differential $\vf_D$.}\label{dual}

Given a Strebel differential $\vf\in \hat{\mathcal S}_{\gg,s}$ consider the following quadratic differential
\be\label{defdual}
\vf_D\,dz^2\equiv \frac{4\,\vf\,e^{2\pi i\,l(z)}}{\left(1-e^{2\pi i\,l(z)}\right)^2}\,dz^2
=-\frac{1}{\pi^2}\left(d\left[\ln\frac{1+e^{\pi i l(z)}}{1-e^{\pi i l(z)}}\right]\right)^2,
\ee where we have defined 
\be
l(z)\equiv \int_{z'}^{z}dz\sqrt{\vf},%,\qquad\sqrt{\vf}dz=dl.
\ee and $z'$ is one of the zeros of the differential $\vf$.  Note that $\vf_D$ has correct
properties under coordinate transformations. The differential is well defined because all the edges
of $\vf$ are integer (and this also implies that  $\vf_D$ has the same periodicity
properties as $\vf$). Let us consider the behavior of $\vf_D$ near special
points of $\vf$. First, near a double pole $z_a$ of $\vf$ we have
\be
\vf(z\to z_a)\sim-\frac{p_a^2}{4\pi^2}\frac{1}{(z-z_a)^2}.
\ee Using this we obtain
\be
e^{2\pi i\,l(z)}\propto e^{- p_a\int^{z}\frac{dz}{z-z_a}}=(z-z_a)^{- p_a}.
\ee The above implies for $\vf_D$ that
\be
\vf_D\propto \frac{(z-z_a)^{-2}(z-z_a)^{-p_a}}{(z-z_a)^{-2p_a}}=(z-z_a)^{p_a-2}.
\ee Thus the differential $\vf_D$ has a zero of valence $p_a-2$ where the differential
$\vf$ has a double pole with circumference $p_a$. Consider now a zero $z_b$ of valence $m_b$
of the differential $\vf$. Near the zero we have
\be
\vf(z\to z_b)=A_b\,(z-z_b)^{m_b}+O((z-z_b)^{m_b+1}),
\ee where $A_b$ is some constant. Remember that all the edges of differential $\vf$ have
integer lengths by assumption and thus
\be
e^{2\pi i\,l(z)}=1-2\pi i\sqrt{A_b}\frac{2}{m_b+2}(z-z_b)^{m_b/2+1}
+O((z-z_b)^{m_b/2+2}).
\ee Thus we obtain for the differential $\vf_D$
\be
\vf_D\sim\frac{4\,A_b\,(z-z_b)^{m_b}}{\left(1-(1-2\pi i\sqrt{A_b}\frac{2}{m_b+2}(z-z_b)^{m_b/2+1})\right)^2}=-\frac{(m_b+2)^2}{4\pi^2}\frac{1}{(z-z_b)^2}.
\ee Thus we conclude that at the positions of the zeros of $\vf$ with
valence $m$ the differential $\vf_D$ has double poles of circumference $m+2$. Moreover,
at the regular points on the critical curve of $\vf$ having integer distance from $z'$ the 
same argument as above implies that the differential $\vf_D$ has a double pole with
circumference $2$. The differential $\vf_D$ has no other poles or zeros. We have shown that the differential $\vf_D$ satisfies all the properties
of the dual quadratic differential from section \ref{punct}, and we just have to prove that it
also is a Strebel differential.

\begin{figure}[htbp]
 \begin{center}
$\begin{array}{c@{\hspace{0.6in\vspace{0.3in}}}c}
\epsfig{file=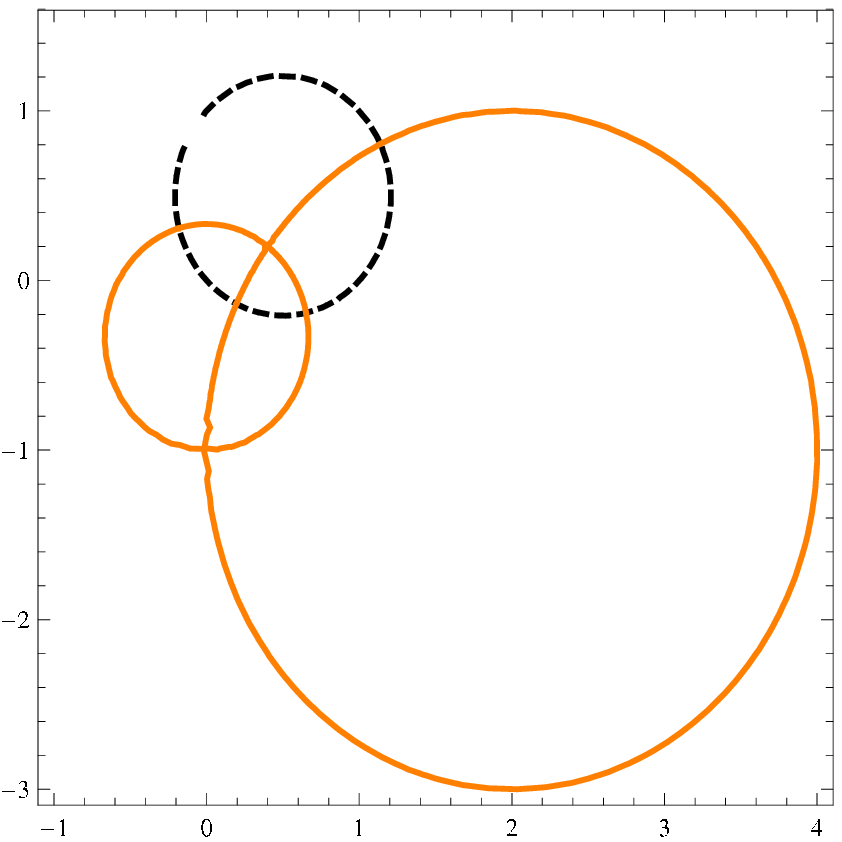,scale=0.6}&\epsfig{file=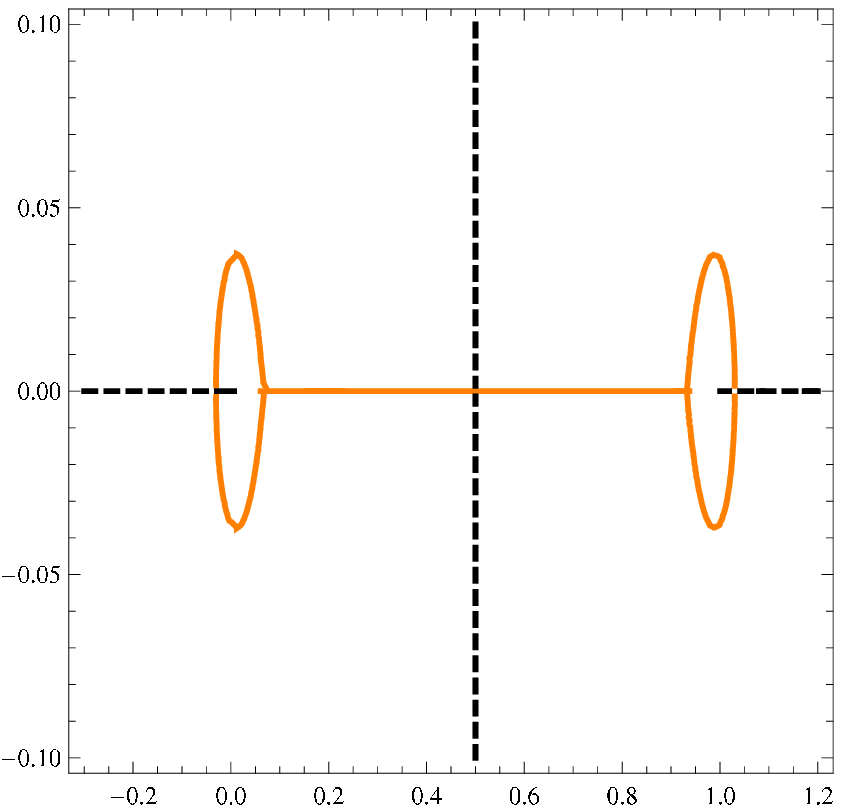,scale=0.6}
\\ [0.2cm]
\end{array}$
\caption{Depicted here are two examples of the critical curves of differentials along with their duals. On the left, the dashed
line corresponds to the Feynman diagram (corresponding to the dual differential $\vf_D$ in our notations) of $\langle\left(TrQ^2\right)^4\rangle$. The insertions are at $z=0,\,1,\,i,\,\frac{1}{5}(2-i)$. 
This is a very simple example as $\vf_D$ has no zeros or simple poles at all and as its critical curve we draw a closed 
horizontal leaf passing through the poles of $\vf$. On the right, the curves corresponding
to the differentials of  $\langle TrQTrQTrQ^4\rangle$ are depicted. The insertions are at $0,\,1,\,$ and $\infty$. The lines of the dual curve (the dashed lines) meet at $z=\infty$, where $\vf_D$ 
has a double zero. 
}\label{DualStrebel}
\end{center}
\end{figure}

\begin{figure}[htbp]
 \begin{center}
$\begin{array}{c@{\hspace{0.6in\vspace{0.3in}}}c}
\epsfig{file=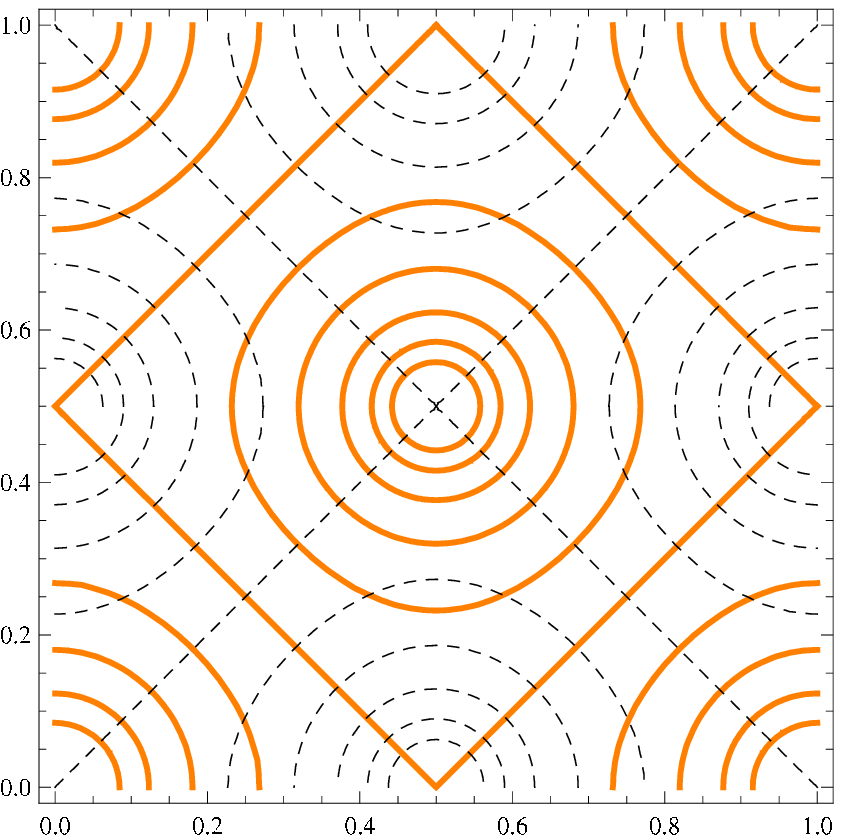,scale=0.6}&\epsfig{file=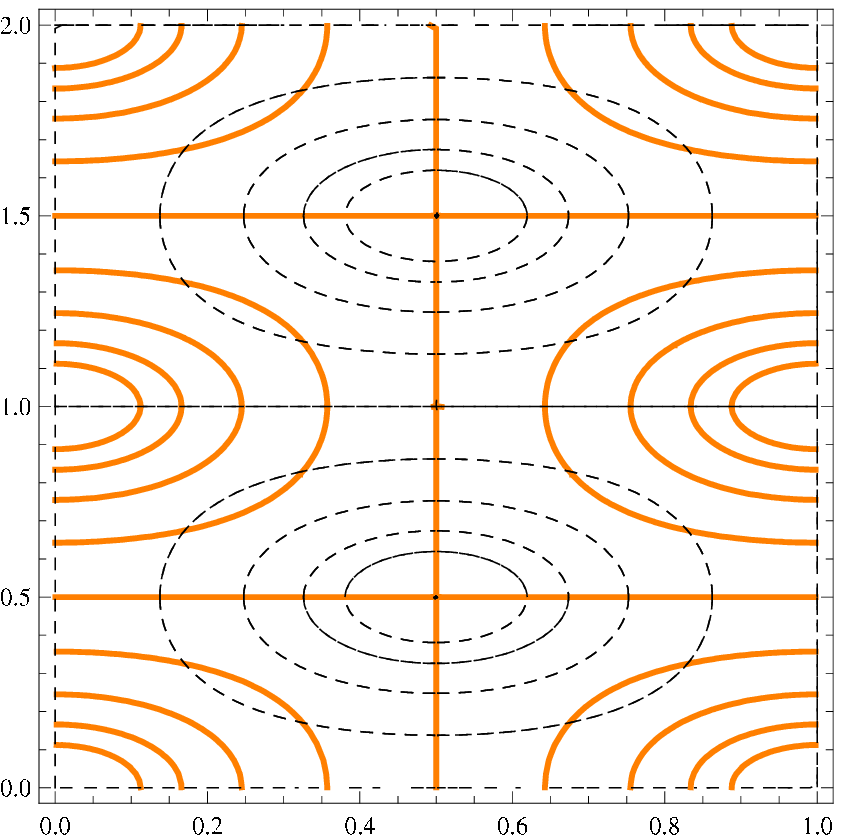,scale=0.6}
\\ [0.2cm]
\end{array}$
\caption{Depicted here are the two diagrams contributing to $\langle TrQ^4TrQ^4\rangle_{\gg=1}$. On the left, we have a diagram without self contractions
(the fundamental domain of the torus is depicted).
This diagram gets a single contribution from the point in moduli space with insertions at $z=0,\,\frac{i+1}{2}$ and with the torus modulus $\tau=i$. 
On the right is the diagram with self contractions. Here the insertions are at $z=0$ and $z=i$ with the torus modulus $\tau=2i$. 
The dashed lines correspond to the horizontal leaves of $\vf_D$  and the solid lines to horizontal leaves of $\vf$.
}\label{DualStrebel2}
\end{center}
\end{figure}

In order to prove that the $\vf_D$ differential is Strebel we have to show that
all the distances between the zeros of $\vf_D$ are real. Essentially, we have to prove that
the following integrals are real
\be
l_D(z_a)\equiv\int_{\tilde z'}^{z_a}dz\sqrt{\vf_D}\in {\mathbb R},
\ee where $\tilde z'$ and $z_a$ are zeros (or simple poles) of $\vf_D$ and thus double poles
of $\vf$ by construction.  
 Using the fact that $l(z\to z_a)\;\to\;i\ln (z-z_a)$, and  \eqref{defdual} we obtain
\be\label{lD}
l_D(z_a)=\frac{1}{\pi i}\left.\left(\ln \left[1+e^{\pi i l}\right]-\ln \left[1-e^{\pi i l}\right]\right)\right|_{l(\tilde z')}^{l(z_a)}
\;\in\;{\mathbb Z}.
\ee 
Thus we have shown that $\vf_D$ is a valid Strebel differential with integer edge lengths. The 
critical curve of $\vf_D$ is the Feynman diagram, the skeleton graph of which
is dual to the critical curve of $\vf$.
Finally, using \eqref{lD} one can  show that the dual differential of $\vf_D$ ($\vf_{DD}$) is $\vf$,
\be
\vf_{DD}&=&\frac{4\,\vf_D\,e^{2\pi i\,l_D(z)}}{\left(1-e^{2\pi i\,l_D(z)}\right)^2}=
\frac{16\,\vf\,e^{2\pi i\,l}}{\left(1-e^{2\pi i\,l}\right)^2}\,
\left[\frac{1+e^{\pi i l}}{1-e^{\pi i l}}\right]^2\frac{1}{\left(1-\left[\frac{1+e^{\pi i l}}{1-e^{\pi i l}}\right]^2\right)^2}=\vf.
\ee As a simple example of the above we consider the following differential ($\wp(z|\tau)$ is the Weierstrass elliptic function\footnote{
See for instance \cite{Aharony:2006th} for examples of Strebel differentials on a torus.})
\be
\vf(z)=-\frac{4}{\pi^2}\wp(z|\tau=i).
\ee We regard this differential as a Strebel differential on a torus with $\tau=2i$. This differential has two double poles of circumference
four and two double zeros in the fundamental domain. We use \eqref{defdual} to find the dual differential 
\be
l(z)=\frac{1}{\pi}\arcsin\wp(z|\tau=i)\;\to\; \vf_D(z)=\frac{4}{\pi^2}\frac{1}{\wp(z|\tau=i)}=-\frac{4}{\pi^2}\wp(z-\frac{1+i}{2}|\tau=i).\nonumber\\
\ee 
Several additional examples of differentials along with their duals are depicted in figures \ref{DualStrebel} and \ref{DualStrebel2}.

\section{Measure details.}\label{measure}

On a given Riemann surface with a (dual Strebel) metric $g_D$ we define a complete set of functions
\be\label{defpsi}
\d\bar\d \psi_n=-\sqrt{g_D} \e_n\psi_n,\qquad&&\int d^2z\sqrt{g_D}\psi_n\psi_m=\delta_{n,m},\\
\sqrt{g_D}\sum_n\psi_n(z)\psi_n(z')&=&\delta^{(2)}(z-z').
\nonumber
\ee
Given a function on a Riemann surface we make the following expansion
\be
X(z,\bar z)=\sum_n a_n\psi_n,
\ee   The zero mode $a_0$ encodes the fact that the scalar is periodic
and takes values in the interval $[0,1]$. The standard path integral measure for a scalar field  $X$ is given by
\be\label{standmeas}
[{\mathcal D}X]_{g_D}=\prod_n da_n.
\ee We want to show that the following path integral measure for the $X$ field can be chosen ($\Delta_{g_D}=\sqrt{\det\,\d\bar\d}$)
\be
[\hat{\mathcal D}X]_{g_D}=\Delta_{g_D}[{\mathcal D}X]_{g_D}\,e^{-\pi\int d^2z\,\d X\bar\d X}.
\ee We have to check that with this definition of the measure  the results of \eqref{stringPv2} are reproduced.
Let us compute the following quantity (denote ${\mathcal H}\equiv Z(\vf_D)\cup P(\vf_D)$),
\be\label{meascond2}
{\mathcal I}&&\equiv\int [\hat{\mathcal D}X]_{g_D} e^{-2\pi i \sum_{k\in {\mathcal H}}P_k X(z'_k)}
\prod_{q=1}^{s+m}\int d^2z_q\,f_{J_q+2}(g,g_D)\,e^{2\pi i J_q X(z_q)}=\\
&&\Delta_{g_D}\prod_{q=1}^{s+m}\int d^2z_q\,f_{J_q+2}(g,g_D)\int [{\mathcal D}X]_{g_D}\,e^{-\pi\int d^2z\,\d X\bar\d X}\, e^{-2\pi i \left[\sum_{k\in {\mathcal H}}P_k X(z'_k)
-\sum_{q=1}^{s+m}J_q X(z_q)\right]}.\nonumber
\ee Here $m$ is the number of puncture operators and $s$ is the number of all the other
operators. In order for the continuous and the discrete definitions of the model to agree the above has to be a sum of products of Kronecker
$\delta$-functions equating the elements of the set of $\{P_k\}$ to the elements of the set $\{J_k\}$. Note also that the $X$ path integral
above looks very similar to a standard correlator of tachyons in bosonic string theory. The difference is that here  the metric (the Strebel metric $g_D$) is singular with some of the insertions sitting at the singularities ($z'_k$s). In the standard string theory the above path integral is regularized by
normal ordering the operators. However, because  some of the operators are inserted at the singularities of the metric this regularization is
not effective in our case and we will employ a slightly different one.
Thus, let us evaluate \eqref{meascond2} directly
using \eqref{defpsi} and \eqref{standmeas}
\be\label{I}
{\mathcal I}
&=&\int \left[\prod_{q=1}^{s+m}d^2z_q\right]\,f_{J_q+2}(g,g_D)\,e^{-\pi\sum_n\frac{\left[\sum_{\mathcal H}P_k\psi_n(z'_k)-\sum_{q=1}^{s+m} J_q\psi_n(z_q)\right]^2}{\e_n}}.
\ee 
Note that the exponential appearing in the integrand is almost always zero. This follows from the fact that $G(z,z')=\sum_n\frac{1}{\e_n}\psi_n(z)\psi_n(z')$ is the Green's function on the Riemann surface and it logarithmically diverges when $z\to z'$.
The only way for the integrand not to vanish is if 
\be\label{cond3}\forall n\; \to\;\sum_{\mathcal H}P_k\psi_n(z'_k)-\sum_{q=1}^{s+m} J_q\psi_n(z_q)=0.\ee
Further we will assume that $f_{J}(g,g_D)$ diverges at the positions of the poles of $\vf_D$ for $J$ corresponding to the puncture operator ($J=0$). A natural (and \textit{diff} invariant)
way to satisfy this assumption is to take $f_{0}\sim\sqrt{g_D}$.
 We also assume
that  $f_{J>0}$ diverges  at the positions of the zeros of $\vf_D$. The natural choice satisfying this is to take 
$f_{J>0}\sim \sqrt{g}$. Note  (see appendix \ref{dual}) that the poles of the metric $g$ and the zeros of $g_D$ are
located at the same positions on the worldsheet.
Assuming the above (and remembering that the number of zeros and simple poles
of $\vf_D$ is equal to $s$) the integrand will not vanish  only  if the set $\{P_k\}$ can be element by element equated to the set $\{J_k\}$ and the set
 $\{z'_k\}$ to be accordingly equated to the set $\{z_q\}$. Thus, we conclude that the integrand
of  ${\mathcal I}$ will be proportional to a sum of products
 of Kronecker $\delta$-functions.

In order to obtain a finite result for ${\mathcal I}$ we have to regularize the integrals in \eqref{I}.
 For instance, we can restrict to a finite number of $\psi_n$ and then the exponential in the integrand will be smeared.
  We can regulate the metrics by cutting small holes around the poles. The exact prescription for the regularization will be encoded in the over-all normalization of the vertex operators.
 Thus finally, we will define the  operators as
\be
\tilde {\mathcal O}_0&=&C_0 \int d^2z\sqrt{g_D}e^{-4\pi i X},\\
{\mathcal O}_{J>0}&=& C_J \int d^2z\,\sqrt{g}\,e^{2\pi i(J-2) X}.\nonumber
\ee 
The constants $C_J$ and $C_0$ are the normalization constants and their value depends on the explicit regularization procedure
used.

\end{document}